# Title: Non-trivial charge-to-spin conversion in ferromagnetic metal/Cu/Al$_2$O$_3$ by orbital transport

**Authors:** Junyeon Kim[1], Dongwook Go[2], Hanshen Tsai[1,3], Daegeun Jo[2], Kouta Kondou[1], Hyun-Woo Lee[2], YoshiChika Otani[1,3]*

**Affiliations:**

[1]Center for Emergent Matter Science, RIKEN, Wako, Saitama 351-0198, Japan

[2]Deparment of Physics, Pohang University of Science and Technology, Pohang 37673, Korea

[3]Institute for Solid State Physics, University of Tokyo, Kashiwa, Chiba 277-8581, Japan

**Abstract:** Efficient spin/charge interconversion is desired to develop innovative spin-based devices. So far, the interconversion has been performed by using heavy atomic elements, strong spin-orbit interaction of which realizes the interconversion through the spin Hall effect and the Edelstein effect. We demonstrate highly efficient charge-to-spin conversion in a ferromagnetic metal/Cu/Al$_2$O$_3$ trilayers, which do not contain any heavy element. The resulting spin torque efficiency is higher than those of conventional spin Hall and Rashba systems consisting of heavy elements such as Pt and Bi. Our experimental results qualitatively deviate from typical behaviors arising from spin transport. However, they are surprisingly consistent with the behaviors arising from the orbital transport. Our results thus demonstrate a new direction for efficient charge-to-spin conversion through the orbital transport.

**One Sentence Summary:** Efficient spin control by orbital torque.

**Main Text:**

The charge-to-spin conversion allows for electrical control of magnetism and is a promising tool to realize spin-based memory and logic devices. The conversion relies on either the spin Hall effect (SHE) or the Edelstein effect (EE), and to make these effects strong, heavy atomic elements with strong spin-orbit interaction (SOI) are indispensable (*1-6*). Heavy elements are problematic for practical device applications, however, since their high electrical resistivity makes the energy-efficiency of such devices poor (*7-10*). Moreover heavy elements are often incompatible with mass production.

Here we report highly efficient charge-to-spin conversion in ferromagnetic metal (FM)/Cu/Al$_2$O$_3$ trilayers that contain no heavy elements. Conversion properties of the trilayers are qualitatively distinct from those of the conventional conversion based on the SHE or the EE, and consistent instead with a novel mechanism based on the orbital transport (*11*). This result provides a way to circumvent problems of heavy elements and widens the material choice for the conversion.

We probe the charge-to-spin conversion in a Co$_{25}$Fe$_{75}$ (CoFe, 5 nm)/Cu (10 nm)/Al$_2$O$_3$ (20 nm) trilayer via the spin torque ferromagnetic resonance (ST-FMR) technique (Fig. 1(a), (b)) (*12*). Each resonance peak in a ST-FMR spectrum can be decomposed into a symmetric component $S$ and an anti-symmetric component $A$ (Fig. 1(c)). When a radiofrequency (rf) charge current is injected into the trilayer, current-induced Oersted field $H_{rf}$ generates $A$ whereas the spin torque (ST) due to the charge-to-spin conversion generates $S$. Thus the efficiency $\theta$ of the conversion is related to the ratio $S/A$ as follows (*12*),

$$\theta = \frac{S}{A}\frac{4\pi M_s e t_{FM} d_{Cu}}{\hbar}\left[1+\frac{4\pi M_s}{H_{ext}}\right]^{1/2}. \qquad (1)$$

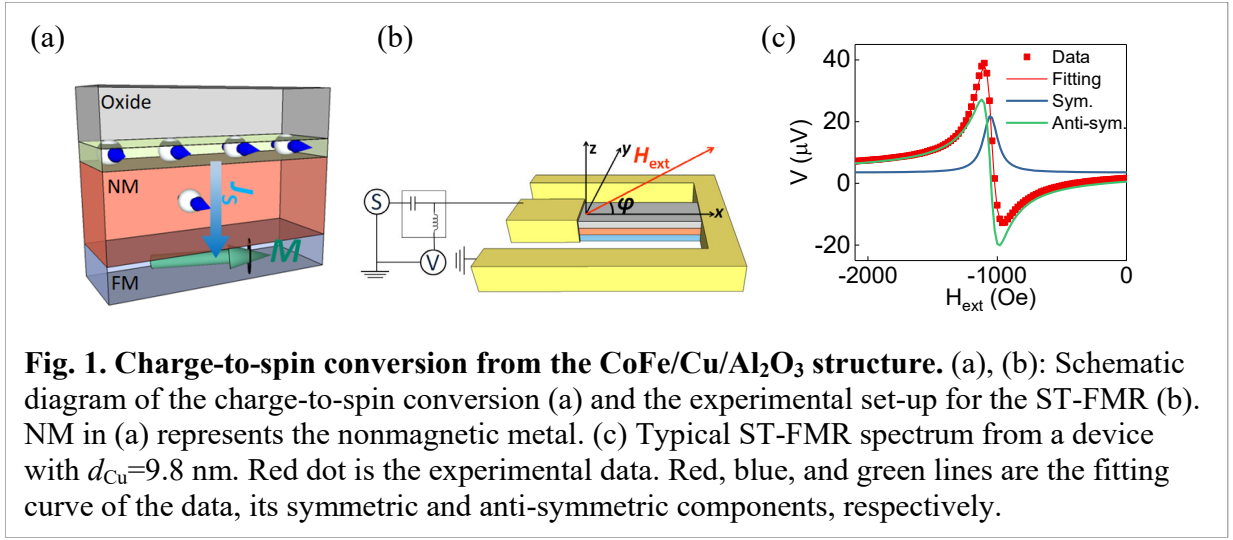

**Fig. 1. Charge-to-spin conversion from the CoFe/Cu/Al$_2$O$_3$ structure.** (a), (b): Schematic diagram of the charge-to-spin conversion (a) and the experimental set-up for the ST-FMR (b). NM in (a) represents the nonmagnetic metal. (c) Typical ST-FMR spectrum from a device with $d_{Cu}$=9.8 nm. Red dot is the experimental data. Red, blue, and green lines are the fitting curve of the data, its symmetric and anti-symmetric components, respectively.

Here, $4\pi M_s$ is the saturation magnetization of the FM, $e$ is the unit charge, $t_{FM}$ is the thickness of the FM layer, $d_{Cu}$ is the thickness of the Cu layer, and $H_{ext}$ is the external field (see Supplementary Materials S2 for the detailed description). For the CoFe/Cu/Al$_2$O$_3$ trilayer, we obtain $\theta \approx 0.13$ which is similar to those for the other layer structures that utilize strong SHE of heavy elements (*3, 13*). This $\theta$ value is thus surprisingly large considering that there is no heavy element in the trilayer.

To find the origin of large $\theta$ without a heavy element, we vary the constituent material of the trilayer. We vary the FM material from CoFe to Fe, Ni$_{80}$Fe$_{20}$ (Py), and Ni, and evaluate $\theta$ for FM ($t_{FM}$ nm)/Cu (0-25 nm)/Al$_2$O$_3$ (20 nm) structures. The thickness of the FM layer ($t_{FM}$) is chosen to 5 nm for FM = Fe and Py, but to a larger value 12 nm for FM=Ni since it has smaller $4\pi M_s$ than the other FMs. We find more than one order of magnitude change in the efficiency $\theta$ with the FM material variation; In contrast to $\theta \approx 0.13$ for CoFe, $\theta$ is essentially zero for Ni, ~0.003 for Py, ~0.1 for Fe (Fig. 2(a)). The value for Py is consistent with that for a similar structure reported earlier (*14*). The variation of $\theta$ with the FM material occurs also in a device that utilizes the SHE or the EE. However the increment of such a variation is a factor of 3 at best (*15*). Thus, the variation of $\theta$ more than one order of magnitude in Fig. 2(a) is anomalously large. We suspect the FM variation of $\theta$ is strongly dependent on an interface quality. Indeed, cross-sectional transmission electron microscope (TEM) images show that for FM=CoFe, FM and Cu atoms are clearly separated (Fig. 2(c)) and form a well-ordered interface whereas for FM=Ni, FM and Cu atoms are mixed (Fig. 2(d)) and form a magnetic dead layer at their interface. The images are consistent with the well-known tendency; Cu atoms tend to get mixed (separated) with Ni (from Fe or Co) atoms at FM/Cu interfaces (*16*). This suggests a possible casual relation between the variation of $\theta$ and interface quality.

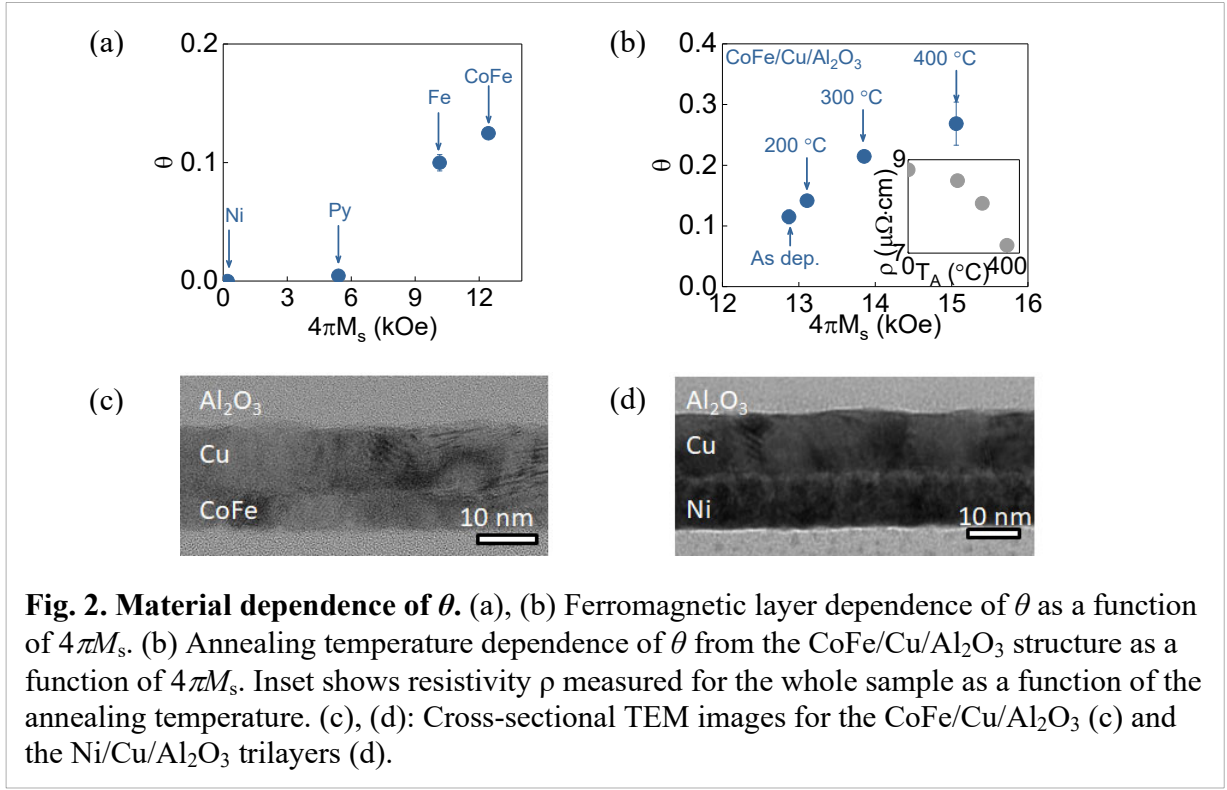

**Fig. 2. Material dependence of $\theta$.** (a), (b) Ferromagnetic layer dependence of $\theta$ as a function of $4\pi M_s$. (b) Annealing temperature dependence of $\theta$ from the CoFe/Cu/Al$_2$O$_3$ structure as a function of $4\pi M_s$. Inset shows resistivity $\rho$ measured for the whole sample as a function of the annealing temperature. (c), (d): Cross-sectional TEM images for the CoFe/Cu/Al$_2$O$_3$ (c) and the Ni/Cu/Al$_2$O$_3$ trilayers (d).

To check if improved interface quality is important for large $\theta$, we fix the FM material to CoFe and vary instead the interface quality by annealing. Depending on whether CoFe (5 nm)/Cu (12 nm)/Al$_2$O$_3$ (20 nm) trilayers are not annealed (as dep.) or annealed at 200 °C, 300 °C, or 400 °C, its conductivity (and also $4\pi M_s$) increases progressively (Fig. 2(b)). Considering that conductivity of a thin film structure is dominated by the diffusive scattering at interfaces, this implies the improvement of interface quality by annealing (*17*). Figure 2(b) shows that $\theta$ grows from ~0.13 to ~0.3 upon annealing. This confirms the importance of the interface quality for large $\theta$ in the CoFe/Cu/Al$_2$O$_3$ trilayer. In contrast, in device structures that utilize the SHE or the EE, a sizable value of $\theta$ can be maintained even with low interface quality that leads to a magnetic dead layer formation (*8, 18*). Ni/Cu/Bi$_2$O$_3$ is also one such example (*19*). Note that $\theta \approx 0.3$ in Fig. 2(b) is close to the value for device structures that utilize the strong SHE of $\beta$-phase W, which has the largest spin Hall angle among single-elemental metallic materials (*7, 9*). Considering that the electrical conductivity of Cu is more than 10 times larger than that of $\beta$-phase W, the CoFe/Cu/Al$_2$O$_3$ device is much more energy-friendly for the ST generation than the CoFeB/$\beta$-phase W structure (*9*).

We find that not only the FM/Cu interface but the Cu/Al$_2$O$_3$ interface is also important for the charge-to-spin conversion. We compare a CoFe (5 nm)/Cu (0~25 nm)/Al$_2$O$_3$ (20 nm) sample with a control sample CoFe (5 nm)/Cu (0~25 nm)/Bi$_2$O$_3$ (20 nm) (*20, 21*), and find that the control samples have systematically smaller values of $\theta$ for a wide range of the Cu layer thickness $d_{Cu}$ despite the presence of the heavy element Bi (Fig. 3(a)). This confirms the importance of the Cu/Al$_2$O$_3$ interface. This raises a possibility that the Cu/Al$_2$O$_3$ interface generates spin through the strong EE and the spin flows through the Cu layer to get injected into the CoFe layer. Considering that Cu is an ideal material for spin transport with a long spin relaxation length (>20 nm) at room temperature (*22*), $\theta$ should then decay slowly with increasing $d_{Cu}$. However, Fig. 3(a) reports more than five-fold decay of $\theta$ upon the $d_{Cu}$ increase from 10 nm to 20 nm. This decay is too drastic and goes beyond possible complications by current shunting. Thus the possibility of the EE at the Cu/Al$_2$O$_3$ interface can be excluded. Another possibility is the SHE due to the Cu layer. Then $\theta$ should saturate with increasing $d_{Cu}$ instead of decaying (*13*). Moreover the spin Hall angle of Cu is two orders of magnitude smaller than that of Pt. Thus the second possibility can also be excluded (*23, 24*).

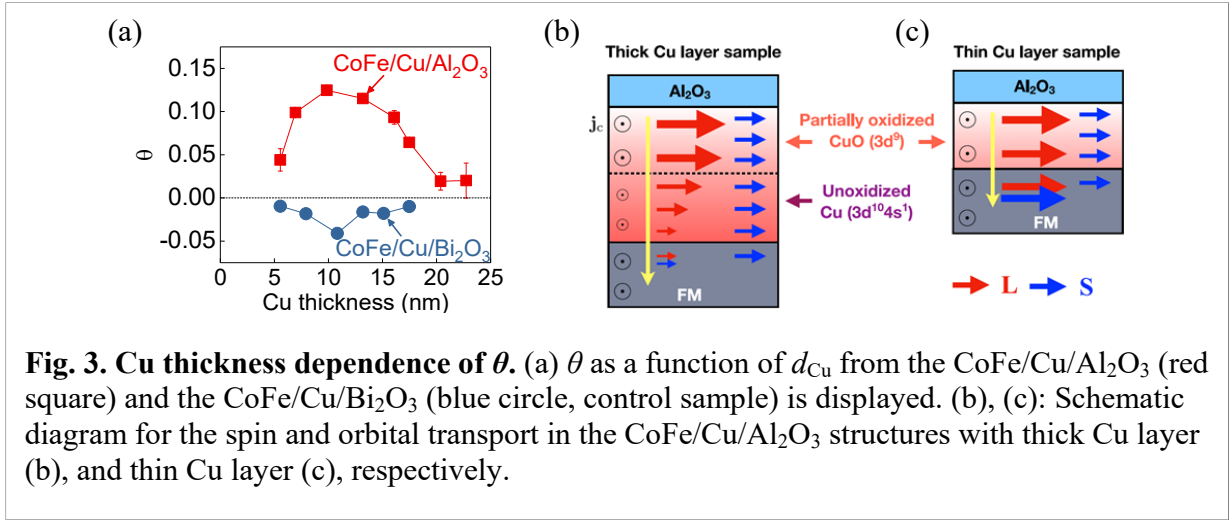

**Fig. 3. Cu thickness dependence of $\theta$.** (a) $\theta$ as a function of $d_{Cu}$ from the CoFe/Cu/Al$_2$O$_3$ (red square) and the CoFe/Cu/Bi$_2$O$_3$ (blue circle, control sample) is displayed. (b), (c): Schematic diagram for the spin and orbital transport in the CoFe/Cu/Al$_2$O$_3$ structures with thick Cu layer (b), and thin Cu layer (c), respectively.

To understand the series of experimental results (Figs. 1,2,3), we note that the SOI is indispensable for the charge-to-spin conversion and that FM elements (Co, Fe, and Ni) have the largest SOI among the atomic elements that constitute the trilayer. Although Cu has slightly larger atomic number than the FM elements, it has mostly *s* character near the Fermi energy, for which the SOI vanishes. Hence the charge-to-spin conversion in our experiment is likely to utilize the SOI of the *FM layer*, implying a need for a new mechanism since conventional SHE- or EE-based mechanisms utilize the SOI of neighboring non-magnetic layers. A recent theory describes such a mechanism, in which (i) the SOI ($\propto \mathbf{L}\cdot\mathbf{S}$) of a FM layer achieves *orbital*-to-spin conversion when (ii) an electrically generated flow of orbital angular momentum $\mathbf{L}$ (charge-to-orbital conversion) is injected into the FM (*11*). According to the estimation in (*11*), a torque (so called *orbital torque*) arising from the injection of such an orbital current can be sizable even when a FM layer has a weak SOI since (iii) an orbital current generated electrically in neighboring nonmagnetic layers can be gigantic (one order of magnitude larger than a similarly generated spin current in Pt) even when their SOI is zero (*23, 25, 26*). Thus in this orbital torque mechanism, gigantic charge-to-orbital conversion in nonmagnetic layers combines with weak orbital-to-spin conversion in a FM layer to generate intermediate but still sizable charge-to-spin conversion. Although the orbital torque mechanism is illustrated in (*11*) with the orbital Hall effect as a means for the charge-to-orbital conversion, the same mechanism applies also when the orbital EE (*27-30*) works as a means for the charge-to-orbital conversion as illustrated in Supplementary Materials S8. Hence the orbital Hall effect and the orbital EE are the orbital counterparts of the SHE and the EE, respectively. In view of the orbital torque mechanism, our experimental results can then be understood as follows. Initially, the charge-to-orbital conversion occurs at the Cu/Al$_2$O$_3$ interface by the orbital EE (Fig. 4(a)). The resulting orbital current flows through the Cu layer and gets injected into a FM layer, where the orbital-to-spin conversion occurs and the orbital torque arises (Fig. 4(b)).

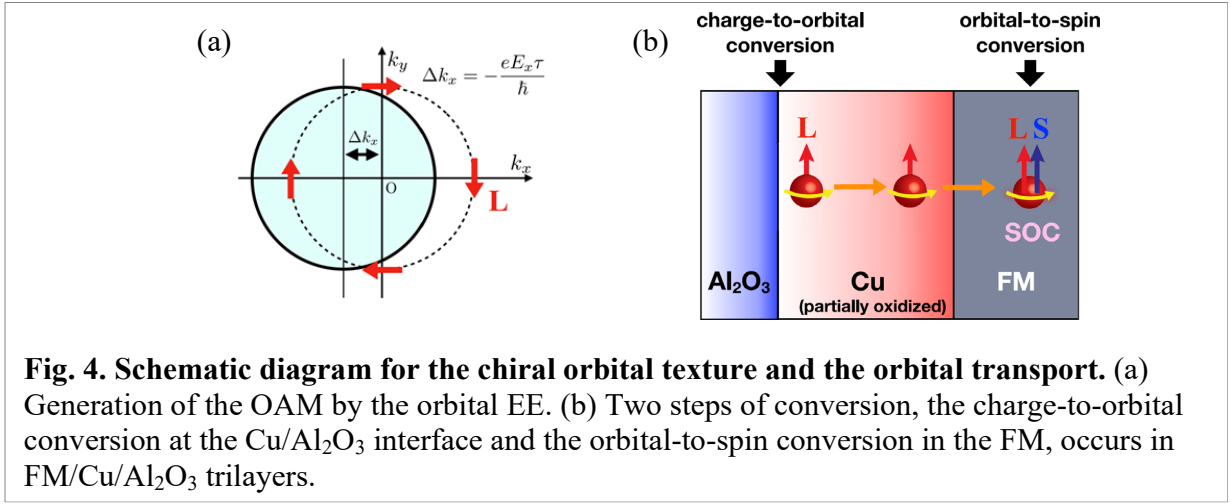

**Fig. 4. Schematic diagram for the chiral orbital texture and the orbital transport.** (a) Generation of the OAM by the orbital EE. (b) Two steps of conversion, the charge-to-orbital conversion at the Cu/Al$_2$O$_3$ interface and the orbital-to-spin conversion in the FM, occurs in FM/Cu/Al$_2$O$_3$ trilayers.

The FM dependence (Fig. 2(a)) and the annealing temperature dependence (Fig. 2(b)) of $\theta$ can be understood from (ii) since the orbital current injection is more vulnerable to and more easily destroyed by the interface disorder than the spin current injection (*11*). $\theta$ being larger for CoFe/Cu/Al$_2$O$_3$ than for CoFe/Cu/Bi$_2$O$_3$ (Fig. 3(a)) can be understood from (iii) if the orbital EE effect is stronger at the Cu/Al$_2$O$_3$ interface than the Cu/Bi$_2$O$_3$ interface. For the $d_{Cu}$ dependence of $\theta$ (Fig. 3(a)), we note on two special properties of Cu. Firstly, whereas Cu is a good material for spin transport (*22*), it is not a good material for orbital transport since its orbital configuration is $3d^{10}4s^1$; partially filled $s$ shell cannot carry **L** and $d$ shell is also hard to carry since it is completely filled. This provides a way to explain the rapid decay of $\theta$ for $d_{Cu} \geq 10$ nm (Fig. 3(b)). Secondly, when Cu is partially oxidized (for instance near the Cu/Al$_2$O$_3$ interface due to the oxygen migration from the Al$_2$O$_3$ layer), its $d$ shell becomes partially filled and contributes to orbital transport efficiently (Fig. 3(c)). If the oxygen migration is limited to a certain distance from the Cu/Al$_2$O$_3$ interface, it provides a way to explain why the $\theta$ decay is absent for small $d_{Cu}$. Moreover considering that the atomic ordering within the Cu layer increases with $d_{Cu}$, the vulnerability of the orbital current to atomic disorder provides an explanation to the increase of $\theta$ for small $d_{Cu}$. One direct implication of the orbital torque mechanism is that for $d_{Cu}$ sufficiently thick, then the orbital current is mostly blocked and only the spin current, which is small since it is generated by weak EE near the Cu/Al$_2$O$_3$ interface, can reach the FM layer (Fig. 3(b)). In this regime of $d_{Cu}$, $\theta$ should not be sensitive to the FM/Cu interface quality. Our measurement for $d_{Cu} \sim 20$ nm verifies this prediction. (see Supplementary Materials S7 and Fig. S11).

**Acknowledgements**

The authors acknowledge D. Hashizume, T. Kikitsu, D. Inoue, and A. Nakao for various discussions on the material characterization. **Funding:** J.K. H.T, K.K and Y.O were supported by a Grant-in-Aid for Scientific Research on the Innovative Area, 'Nano Spin Conversion Science' (Grant No. 26103002). D.G. and H.-W.L acknowledge the financial support of the SSTF (Grant No. BA-1501-07). **Authors contributions:** J.K. conceived the study, and J.K. designed the study with the help of H.-W.L and Y.O. J.K. fabricated the devices. J.K. measured the spin transport properties with the assistance of K.K. J.K and H.T. carried out the material characterization. D.G. and H.-W. L. provides a theoretical assertion. J.K, D.G, and H.-W.L wrote the manuscript. All authors contributed to the analysis and editing. **Competing interests:** The authors declare no competing financial interests. **Data and materials availability:** All data is available in the manuscript or supplementary materials.

**Supplementary Materials:**

Materials and Methods

Figures S1-S13

References (*31-40*)